\documentclass[acmsmall,screen,nonacm]{acmart}
\AtBeginDocument{%
  }

\usepackage{graphicx}
\usepackage{subcaption}
\usepackage{xcolor}
\usepackage{multirow}
\usepackage[most]{tcolorbox}
\usepackage{adjustbox}
\usepackage[table]{xcolor}
\usepackage{quantikz}
\usepackage{tikz}
\usetikzlibrary{arrows.meta, positioning, calc}
\definecolor{stagefill}{RGB}{226,236,247}
\definecolor{edgecol}{RGB}{40,63,90}
\definecolor{consA}{RGB}{0,114,178}
\definecolor{consB}{RGB}{0,158,115}
\definecolor{consC}{RGB}{230,159,0}
\definecolor{defred}{RGB}{213,94,0}
\tikzset{
  stage/.style={draw=edgecol, line width=0.8pt, fill=stagefill, rounded corners=2pt, minimum height=10mm, minimum width=17mm, align=center},
  inputbox/.style={draw=edgecol, line width=0.8pt, rounded corners=3mm, minimum height=10mm, minimum width=17mm, align=center},
  artifact/.style={draw=edgecol, line width=0.8pt, fill=black!8, minimum height=8mm, minimum width=17mm, align=center},
  rbox/.style={draw=edgecol, line width=0.8pt, fill=black!8, minimum height=7mm, minimum width=14mm, align=center},
  ghost/.style={draw=black!45, dashed, minimum height=10mm, minimum width=15mm, align=center, text=black!60},
  flow/.style={-{Stealth[length=2.6mm]}, line width=0.9pt},
  defect/.style={font=\scriptsize\bfseries, text=defred},
  trust/.style={circle, inner sep=0pt, minimum size=2.2mm},
  span/.style={{Bar[width=2.2mm]}-{Stealth[length=2.6mm]}, line width=0.9pt},
}
\usepackage{algorithm}
\usepackage{algpseudocode}
\usepackage{amsmath}
\newcommand{\Inconsistent}{\textsc{Inconsistent}}
\newcommand{\Consistent}{\textsc{Consistent}}
\newcommand{\Gather}{\textsc{Gather}}
\newcommand{\dom}{\operatorname{dom}}
\tcbset{
  colback=blue!5!white,    
  colframe=blue!40!black,  
  boxrule=0.5pt,           
  arc=2mm,                 
  left=1mm, right=1mm, top=1mm, bottom=1mm,  
  enhanced
}

\setcopyright{acmlicensed}
\copyrightyear{2018}
\acmYear{2018}
\acmDOI{XXXXXXX.XXXXXXX}
\acmConference[Conference acronym 'XX]{Make sure to enter the correct
  conference title from your rights confirmation email}{June 03--05,
  2018}{Woodstock, NY}
\acmISBN{978-1-4503-XXXX-X/2018/06}

\begin{document}

\title{A Testing Method for Detector Error Models in Quantum Error Correction Compilers}

\author{Junjie Luo}
\orcid{0009-0008-7821-6879}
\affiliation{%
  \institution{RIKEN Center for Quantum Computing}
  \city{Wako}
  \state{Saitama}
  \country{Japan}
}
\email{junjie.luo@riken.jp}

\author{Yasunari Suzuki}
\orcid{0000-0002-8005-357X}
\affiliation{%
  \institution{RIKEN Center for Quantum Computing}
  \city{Wako}
  \state{Saitama}
  \country{Japan}}
\email{yasunari.suzuki@riken.jp}

\renewcommand{\shortauthors}{Luo and Suzuki}

\begin{abstract}

A quantum error correction (QEC) compiler produces, from a code and a noise model, both the physical circuit to execute and a detector error model (DEM). The DEM lists the probability of each error mechanism and the measurements it flips. Decoders consume the DEM without verification, so a defect that describes the noise incorrectly calibrates the decoder against wrong rates and silently degrades the logical error rate. Existing techniques for testing quantum software judge correctness by crashes or by statistics of sampled outputs. DEM defects that only distort rates lie below what sampling can resolve at the error rates where QEC operates. We present DEMVal, a testing method that validates each model the compiler emits against a reconstructed reference, with rates obtained from outside the compiler under test. It compares the two models exactly, so its sensitivity is independent of the physical error rate. The reference can be built in three ways that jointly cover the pipeline from source program to emitted model. Applied to NVIDIA's CUDA-Q QEC, Riverlane's Deltakit, and the deq component of Microsoft's QDK-EC, DEMVal found six defects, all confirmed by the developers and fixed. Sampling could have exposed only the four defects that drop, merge, or misplace mechanisms, and exposing all four requires millions of samples per experiment. Even analytically optimal moment tests miss the two rate-level defects, while exact model comparison detects all six with no samples. To our knowledge, DEMVal is the first systematic testing method for the error-model semantics of QEC compilers.

\end{abstract}

\begin{CCSXML}
<ccs2012>
   <concept>
       <concept_id>10011007.10011006.10011041</concept_id>
       <concept_desc>Software and its engineering~Compilers</concept_desc>
       <concept_significance>500</concept_significance>
       </concept>
   <concept>
       <concept_id>10010583.10010786.10010813.10011726.10011728</concept_id>
       <concept_desc>Hardware~Quantum error correction and fault tolerance</concept_desc>
       <concept_significance>500</concept_significance>
       </concept>
   <concept>
       <concept_id>10011007.10011074.10011099.10011102.10011103</concept_id>
       <concept_desc>Software and its engineering~Software testing and debugging</concept_desc>
       <concept_significance>500</concept_significance>
       </concept>
 </ccs2012>
\end{CCSXML}

\ccsdesc[500]{Software and its engineering~Compilers}
\ccsdesc[500]{Hardware~Quantum error correction and fault tolerance}
\ccsdesc[500]{Software and its engineering~Software testing and debugging}


\maketitle

\section{Introduction}
\label{sec:intro}

Quantum computing has promising applications in areas such as quantum chemistry~\cite{QCchem} and cryptography~\cite{cryptography}. However, realizing these applications requires running large-scale quantum computations on real hardware, and real quantum devices are noisy. Without correction, this noise builds up over a computation and corrupts its result, so running such applications reliably at scale requires fault-tolerant quantum computing (FTQC)~\cite{FTQC, FTQC2}. FTQC is built on quantum error correction (QEC)~\cite{QEC, Steane}, which encodes each logical qubit, the qubit an algorithm computes on, redundantly across many of the machine's noisy physical qubits. This redundancy lets errors be detected and corrected, so the computation can reach the accuracy its application demands.

Achieving this in practice is a compilation problem, and it falls to a QEC compiler. Several such compilers are now in wide use, such as NVIDIA's CUDA-Q QEC~\cite{cudaqQEC}, Riverlane's Deltakit~\cite{deltakit}, and Microsoft's QDK-EC~\cite{Microsoft_QDK_EC}. From an error-correcting code, a noise model, and an experiment to run, the compiler produces a physical circuit that the hardware executes. The circuit's measurements reveal where errors are likely to have occurred, and a classical algorithm called a decoder reads them to infer which errors happened and how to correct them. To do so, the decoder needs a model of how errors appear in the measurements. The compiler provides this model as a detector error model (DEM), which lists, for each error, the measurements it flips and with what probability. The DEM is the decoder's only information about the circuit's noise. 

Like any large piece of software, QEC compilers have bugs. A defect here is especially serious because the decoder uses this model without verification. When a compiler emits a DEM that describes the circuit incorrectly, the decoder is calibrated against the wrong noise, and the logical error rate (LER) silently degrades. Testing gate-level quantum compilers has received considerable attention, for platforms such as Qiskit~\cite{qiskit2024} and Q\#~\cite{Svore_2018} and through methods such as QDiff~\cite{QDiff} and MorphQ~\cite{MorphQ}. QEC compilers, in contrast, have gone largely untested, because testing them faces two difficulties.

The first difficulty is that the methods used to test gate-level compilers do not carry over. These methods usually check a compiler by comparing the quantum state a program produces, or by sampling its output distribution. A QEC compiler instead emits a circuit and a DEM, not a quantum state, so there is no state to compare. Sampling the output distribution is still possible, but at the low physical error rates where QEC operates, a miscompilation can shift the measurement statistics by less than the sampling noise. The faulty and correct distributions then become indistinguishable, and the defect goes unnoticed. Our evaluation shows that the defects that only distort rates lie in this range.

The second difficulty is that the metamorphic relations these methods rely on no longer hold. A typical relation replaces a gate, or a subcircuit, with a logically equivalent one and expects the output to be unchanged. In fault-tolerant computing this breaks down, because two logically equivalent gates can carry different amounts of noise. Once the noise differs, the transformed circuit is no longer equivalent to the original, and the metamorphic check has nothing left to compare.

To address these two difficulties, we present DEMVal, a testing method that validates each model a QEC compiler emits against a reconstructed reference. DEMVal generates a corpus of memory experiments, the standard QEC benchmark in which a logical qubit is stored and then read back. The corpus varies the code family, the code distance, the number of rounds, the noise channel, and the physical error rate. For each experiment it obtains the model the compiler emits, and it reconstructs a reference model of the same experiment. The rates of the reference are obtained from outside the compiler under test, from the documented noise semantics of Stim~\cite{stim} or from the published definition of the noise model. DEMVal then compares the two models exactly. It matches errors by the measurements they flip, and the rates of matched errors must agree up to a numerical tolerance. Because the comparison is exact rather than sampled, its sensitivity is independent of the physical error rate. A divergence indicates a defect in how the compiler builds, aggregates, or serializes its model.

The reference can be constructed in three ways, and the choice determines what the comparison can test. Construction A rebuilds the model from the same code, noise model, and experiment that the compiler takes as input. Construction B cross-checks the two artifacts that a single compilation emits against each other. Construction C re-applies a published noise model to the circuit the compiler itself exported. Each construction relies on a different input, and everything between that input and the two models falls inside its testing scope. The six defects we find fall in complementary scopes, which turns the choice of construction into a design dimension of the method rather than an implementation detail.

Applied to three widely used QEC compilers, NVIDIA's CUDA-Q QEC, Riverlane's Deltakit, and the deq component of Microsoft's QDK-EC, DEMVal uncovers six silent semantic defects, three of which yield models that are wrong under any noise semantics. All six were reported, confirmed by the developers, and fixed or merged by the maintainers. The consistent experiments match the reference at machine precision, and every inconsistent experiment exhibits one of the six confirmed defects. We also evaluate the statistical oracles of existing quantum software testing on these six defects. Sampling could have exposed only the four defects that drop, merge, or misplace mechanisms, and exposing all four requires millions of samples per experiment. Even the analytically optimal first- and second-moment tests leave the two rate-level defects undetected at realistic error rates, while all six lie far above our exact oracle's tolerance. To the best of our knowledge, DEMVal is the first systematic testing method for the error-model semantics of QEC compilers.

The main contributions of this work are: 
\begin{itemize}
    \item DEMVal, a testing method that validates the compiler's detector error model against a reference built in three constructions with complementary testing scopes.
    \item An empirical characterization of the sampling oracle class, showing that even analytically optimal moment tests do not detect rate-level defects at realistic error rates.
    \item An implementation of the framework for CUDA-Q QEC, Deltakit, and QDK-EC's deq, uncovering six unique silent real-world defects, all confirmed by the developers and fixed.
\end{itemize}

The rest of the paper is organized as follows. Section~\ref{sec:bg} provides background information relevant to our work. Section~\ref{sec:motiv} presents one of the defects as a motivating example. Section~\ref{sec:md} describes the design of DEMVal. Section~\ref{sec:eval} presents our experimental setup and results. We review related work in Section~\ref{sec:rw}, and conclude the paper in Section~\ref{sec:con}.

\section{Background}
\label{sec:bg}

This section reviews the background underlying our method: quantum-computing basics (Section~\ref{sec:bg-basics}), fault-tolerant quantum computing (Section~\ref{sec:bg-ftqc}), and the QEC compilation flow that emits the detector error model~(DEM), the artifact under test (Section~\ref{sec:bg-dem}).

\subsection{The Basics of Quantum Computing}\label{sec:bg-basics}

Quantum computing is a computational paradigm that exploits features of quantum mechanics to process information. Whereas a classical computer manipulates binary bits using deterministic logic, a quantum computer represents and executes computation as a circuit acting on qubits.

\paragraph{Qubits and Quantum Gates.} The elementary carrier of quantum information is the \textit{qubit}. Rather than being restricted to either 0 or 1, a qubit may occupy a superposition $\ket{\psi}=\alpha\ket{0}+\beta\ket{1}$, where $\alpha$ and $\beta$ are complex amplitudes satisfying $|\alpha|^2+|\beta|^2=1$. An $n$-qubit state is a superposition of $2^n$ computational-basis states $\ket{s_1,s_2,\ldots,s_n}$, where $s_i\in\{0,1\}$, and can be represented by $2^n$ complex amplitudes as $\ket{\psi}=\sum_{s\in\{0,1\}^n}\alpha_s\ket{s}$.

A typical quantum program uses three classes of operations: initialization, unitary gates, and projective measurements. Initialization resets a specified qubit to a target state. A unitary gate deterministically maps $\ket{\psi}\mapsto U\ket{\psi}$, where $U$ is a unitary matrix. A projective measurement is specified by a set of projector matrices $\{\Pi_s\}$. Outcome $s$ occurs with probability $\|\Pi_s\ket{\psi}\|^2$, projects the state to $\Pi_s\ket{\psi}/\|\Pi_s\ket{\psi}\|$, and returns the classical value $s$.

Although these operations form a universal set of quantum computation, we also introduce Pauli gates, Clifford gates, and Pauli measurements because they play a central role in quantum error correction. The matrices
$
I=\begin{pmatrix}1&0\\0&1\end{pmatrix},
X=\begin{pmatrix}0&1\\1&0\end{pmatrix},
Y=\begin{pmatrix}0&-i\\i&0\end{pmatrix},
Z=\begin{pmatrix}1&0\\0&-1\end{pmatrix}
$
and their tensor products, including phase factors, form the Pauli group $\mathcal{P}=\{\pm1,\pm i\}\times\{I,X,Y,Z\}^{\otimes n}$. Unitary gates represented by its elements are Pauli gates. A unitary gate $U$ is a Clifford gate if $UPU^{\dagger}\in\mathcal{P}$ for every $P\in\mathcal{P}$.
A Pauli measurement is a projective measurement with $\Pi_0=(I+P)/2$ and $\Pi_1=(I-P)/2$. It deterministically outputs 0 without changing the state when the state is a $+1$ eigenstate of $P$.

\paragraph{Quantum Programs.} A quantum program is represented as a circuit, i.e., an ordered sequence of operations applied to an initial state. Values sampled during execution are called {\it logical observables} and may control program branches or represent computational results. Typical qubit platforms support only subsets of initialization, unitary gates, and projective measurements; operations supported by the target platform are called physical gates.

\paragraph{Noise Model.} Qubits are sensitive to environmental noise, so physical gates deviate from their ideal actions. Although a general description of noisy quantum gates can be complicated, an appropriate randomization procedure called twirling can reduce their effective noise to probabilistic Pauli insertions. Theoretical frameworks therefore often model qubit noise by inserting Pauli gates drawn from specified distributions before or after ideal quantum gates. We call each Pauli gate that may be drawn from such a distribution an {\it error mechanism}. A $k$-qubit uniform depolarizing channel is a simple example. It selects uniformly from the $4^k-1$ nonidentity Paulis in $\{I,X,Y,Z\}^{\otimes k}\setminus\{I^{\otimes k}\}$, applying each with probability $p/(4^k-1)$, where $0\leq p\leq1-1/4^k$ and $p$ is the depolarizing rate.

The appropriate noise model depends on the qubit platform. A common model applies $k$-qubit uniform depolarizing noise after every $k$-qubit physical operation and during idling. Several named noise models have also been proposed to describe error distributions inspired by actual qubit devices. They use explicit rule sets that express the error rate of each operation as a multiple of one overall parameter. Examples include the standard depolarizing model SD6 and the superconducting-inspired model SI1000~\cite{HoneycombMemory}.

\subsection{Fault-Tolerant Quantum Computing}\label{sec:bg-ftqc}

\paragraph{Stabilizer Codes.}
Current qubit error rates are high and limit the achievable depth of computation under realistic noise. Fault-tolerant quantum computing goes beyond this regime by encoding logical information in the code space of quantum error-correcting~(QEC) codes. This enables detection and correction of inserted Pauli errors with high probability as computation proceeds.

Stabilizer codes define a $k$-qubit code space in $n$ noisy qubits as the simultaneous $+1$ eigenspace of $n-k$ independent commuting $n$-qubit Pauli operators called stabilizer operators. Measuring these operators to check whether a state remains in the code space is called syndrome measurement. If every error mechanism acting on fewer than $d$ qubits either leaves the encoded state unchanged or can be detected by noiseless syndrome measurements, the code has distance $d$ and is denoted $[[n,k,d]]$.

\paragraph{Code Families.}
Several QEC-code families are widely studied. The repetition code~\cite{repetition}, the simplest example, places physical qubits on a line and uses checks that compare neighboring pairs. It protects against either bit-flip~(Pauli-$X$) or phase-flip~(Pauli-$Z$) errors, but not both, making its behavior easy to analyze and making it a common baseline. The surface code~\cite{surfacecode} arranges data and check qubits on a two-dimensional lattice and uses only geometrically local checks. This locality suits planar hardware and has made the surface code a leading candidate for large-scale devices. Its planar layout has unrotated and rotated variants; the rotated variant achieves the same distance with roughly half as many physical qubits. The Steane code~\cite{Steane} encodes one logical qubit in seven physical qubits using checks derived from the classical Hamming code. Other families include the $[[4,2,2]]$ error-detecting code~\cite{422}, Bacon--Shor codes~\cite{Bacon-Shor}, and Floquet codes~\cite{Floquet}, whose checks vary between rounds.

\paragraph{Fault Tolerance and Detector Error Model.}
The usual error-correction guarantee based on code distance assumes that errors occur between noiseless syndrome measurements. In practice, Pauli errors can also occur during syndrome measurements and logical operations and may propagate to multiple qubits. Large code distance alone therefore does not guarantee fault tolerance. Syndrome measurements and encoded operations must be carefully constructed from noisy physical operations so that Pauli errors can be efficiently estimated and tracked.

Several methods have been explored to establish fault-tolerant implementations for given operations, noise models, and stabilizer codes. A common framework for constructing and analyzing such implementations uses the detector error model~(DEM). Logical operations are implemented with physical Clifford gates, Pauli measurements, and state initialization. We call the outcomes of physical Pauli measurements {\it records} and the output quantities of target quantum programs {\it logical observables}. Logical Pauli measurements are designed so that each observable is the XOR of selected records. A {\it detector} is an XOR of records whose value is deterministic in a noiseless execution. Each logical operation is therefore described by a physical-operation sequence, record-to-detector maps, and record-to-observable maps.

Thanks to the properties of Clifford gates and Pauli measurements, the effect of any inserted Pauli error can be computed exactly and efficiently without simulating quantum amplitudes~\cite{HeisenbergRepresentation}. Also, every Pauli error deterministically flips a subset of records and, consequently, subsets of detectors and logical observables. The probability distribution of error mechanisms and their associated flipped detectors and observables form the DEM. Error estimation then infers the flipped observables from the DEM and observed flipped detectors. Most popular error-estimation libraries are implemented based on this formalism~\cite{PyMatching}.

\subsection{QEC Compilation}\label{sec:bg-dem}

\begin{figure}[htbp]
    \centering
    \begin{subfigure}[b]{0.55\textwidth}
        \includegraphics[width=\linewidth]{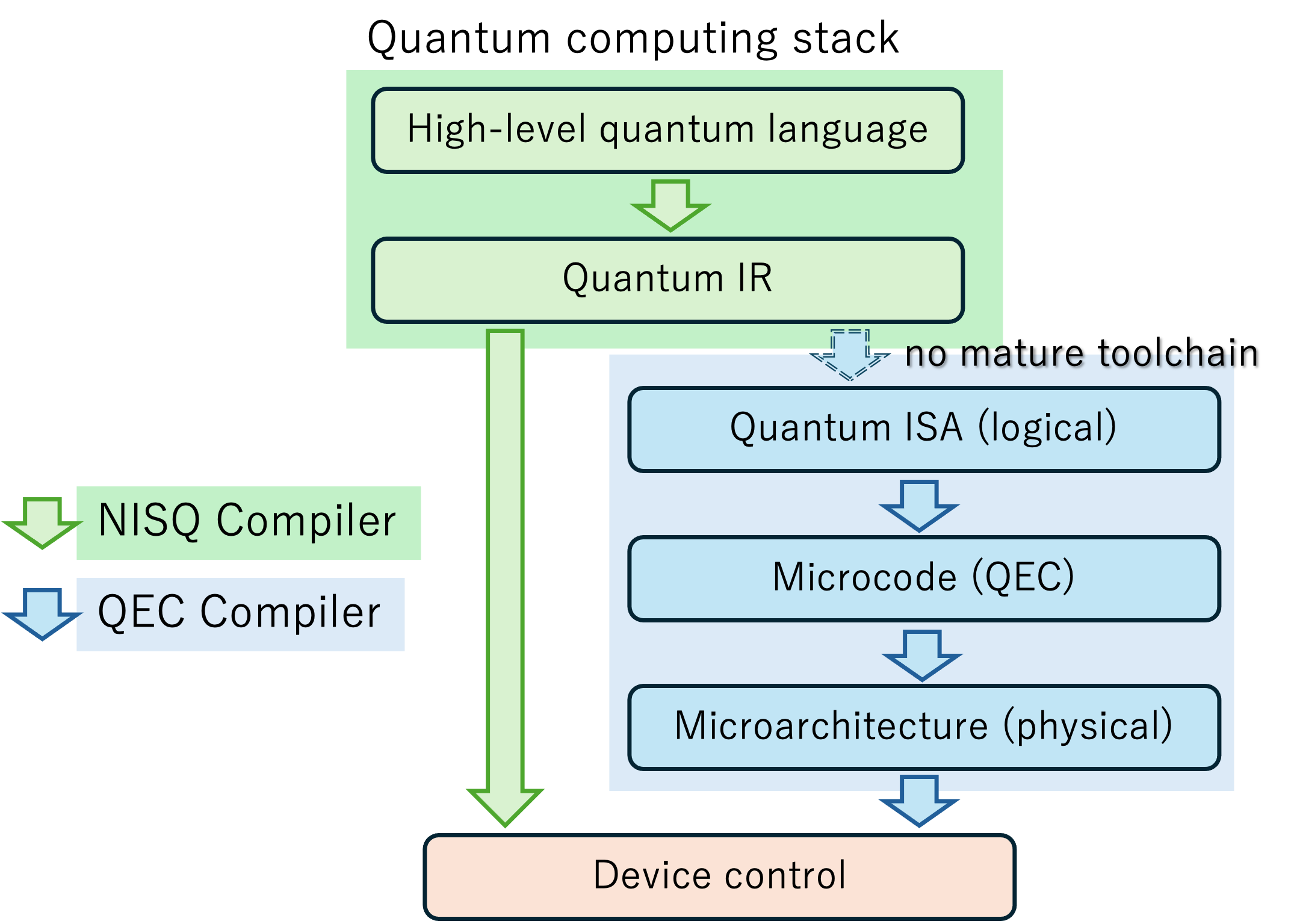}
        \caption{The stack for quantum computing. }
        \label{fig:qc_stack}
    \end{subfigure}\qquad
    \begin{subfigure}[b]{0.35\textwidth}
        \includegraphics[width=\linewidth]{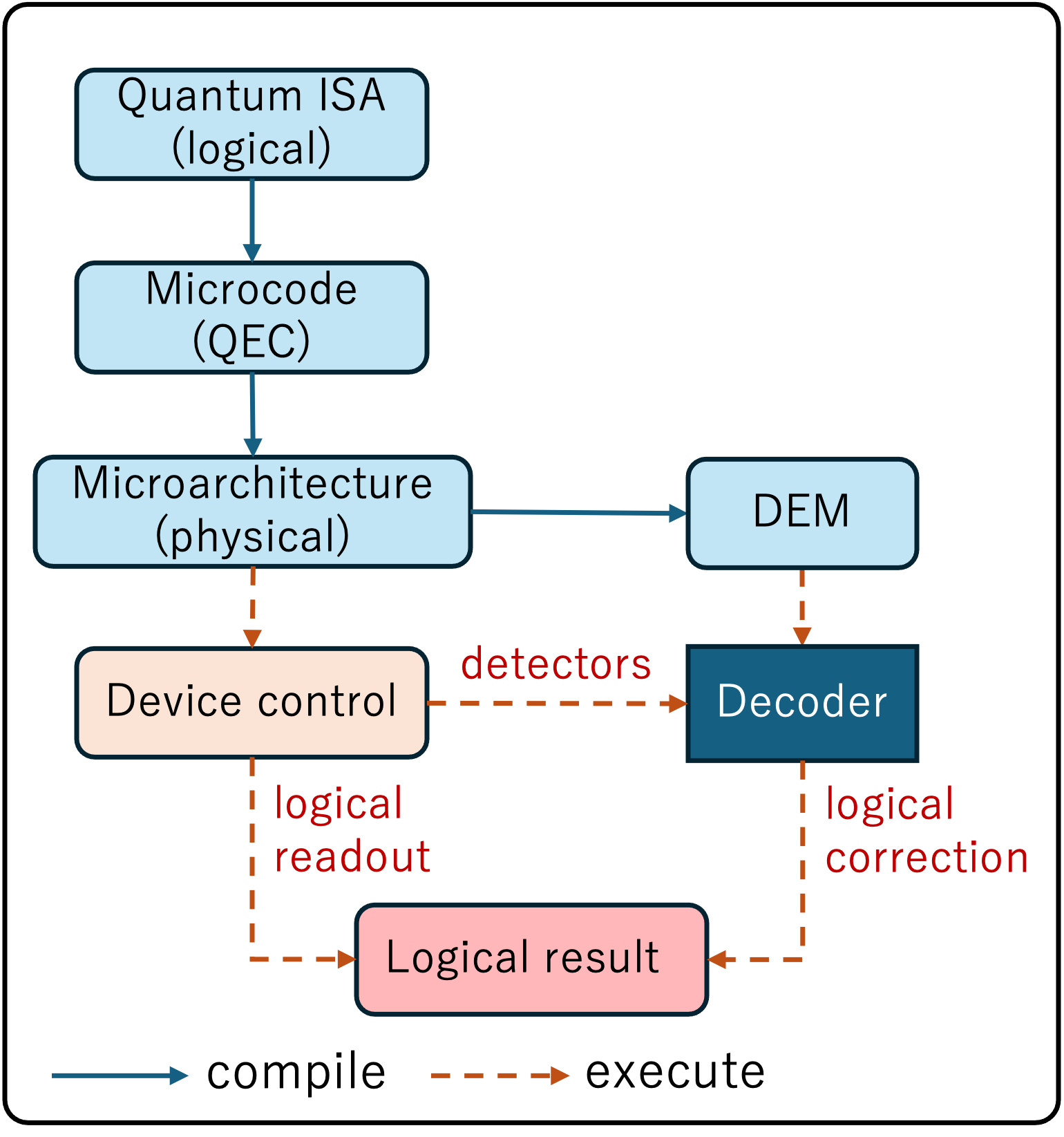}
        \caption{The workflow in QEC compiler and runtime. }
        \label{fig:qec_compiler}
    \end{subfigure}
    \caption{The quantum computing stack and the QEC compilation workflow. (a) The layered stack, adapted from Figure 1 of~\cite{ScalePQA}. A NISQ compiler lowers programs directly to device control, whereas the fault-tolerant flow lowers them through the error-correction layers. (b) The compile-time and run-time data flow of a QEC compiler. The compiler emits the physical circuit together with the detector error model, and at runtime, the decoder consumes the model to turn observed syndromes into logical corrections. The runtime loop on the right lies beyond the boundary our method tests.}
    \label{fig:qc_stack_compiler}
\end{figure}

\paragraph{QEC Compilation Workflow.}
Figure~\ref{fig:qc_stack_compiler} places the components in the software stack for quantum computing. A NISQ compiler lowers a quantum program directly to physical operations. In fault-tolerant quantum computing, available logical operations are determined by the selected QEC code, and a compiler for fault-tolerant quantum computing therefore first lowers the program to logical operations.
At compile time, a DEM must also be generated for the logical-operation sequence. This process requires an assumed noise model in addition to the sequence itself. We call compilers responsible for this process QEC compilers.

Lowering a general quantum IR to this logical layer remains an active research area supported mainly by prototype compilers; no production toolchain yet exists. Existing QEC compilers are therefore invoked directly at the error-correction layer with a QEC code, noise model, and logical-operation sequence. Alongside the physical circuit, the QEC compiler emits the DEM introduced above. At run time, the two artifacts serve different roles: device control executes the physical circuit and streams measurement outcomes, while the decoder reads the DEM, infers errors consistent with observed detectors, and applies corrections to the logical observable. The DEM is generated once at compile time and then consumed without verification. This paper studies the workflow from the input specification to the emitted model.

\paragraph{Description of Detector Error Model.}
A DEM is a list of error mechanisms, each annotated with a probability, the set of detectors it flips, and the set of logical observables it flips. Mechanisms are independent, and two mechanisms with identical effects and probabilities $p$ and $q$ combine with probability $p+q-2pq$. Stim, the de facto standard tool, defines a textual DEM format and computes the exact DEM of a noisy stabilizer circuit~\cite{stim}. Each mechanism occupies one line, with \texttt{D}$k$ denoting detector $k$ and \texttt{L}$k$ denoting logical observable $k$. For example, \texttt{error(0.001) D0 D1 L0} describes a mechanism with probability $10^{-3}$ that flips detectors 0 and 1 and logical observable 0. A complete model is a list of such lines. The compilers tested here emit their models in this format for this purpose.

\section{A Motivating Example}
\label{sec:motiv}

Consider the defect qdk-ec\#68. deq compiles a memory experiment whose noise contains a depolarizing channel at rate $p$. Compilation must express the channel as DEM mechanisms, and so must convert the channel parameter into per-mechanism priors. The two probabilities have different semantics. In the documented channel, exactly one error occurs, and it is $X$, $Y$, or $Z$ with probability $p/3$ each. In a DEM, each mechanism is an independent event that occurs with its own prior $q$. Two may occur together, and their Paulis then multiply, so an $X$ and a $Z$ on the same qubit amount to a $Y$. The priors that reproduce the channel therefore satisfy
\[
q\,(1-q)=\frac{p}{3}, \qquad q=\frac{1-\sqrt{1-4p/3}}{2}=\frac{p}{3}+\Bigl(\frac{p}{3}\Bigr)^{2}+\cdots.
\]
The released compiler instead set $q=p/3$, the channel's own weight and the first term of the correct solution. The mistake is easy to make, and every affected rate in the emitted model is too small by a relative deviation of about $p/3$.

The defect causes no crash, and the decoder uses the model without verification, so the only consequence is a decoder calibrated against wrong rates and a logical error rate above what the code can achieve. At the realistic rate $p=10^{-3}$, the deviation moves every observable syndrome statistic by so little that even the analytically optimal tests on first and second moments detect it with probability near zero at $10^{7}$ samples. Only at artificially elevated rates, such as $p=0.15$, can sampling detect the deviation, at more than $20\sigma$ at $10^{6}$ samples. The defect was nevertheless present in released versions; the maintainers confirmed it and accepted our fix.

DEMVal finds the defect as follows. It rebuilds a reference model of the same experiment from the specification alone, and it obtains the rates of the reference from Stim's documented implementation of the channel. It then compares the emitted model against the reference exactly. On every affected rate the comparison reports the missing terms as a deviation, up to 467 times the $10^{-6}$ tolerance deq itself declares for model comparisons at $p=10^{-3}$, and reaching $5.3\times10^{-2}$ at $p=0.15$. The deviation persists at every error rate, including the realistic ones where sampling cannot detect it. The rest of the paper develops this into a general method.

\section{Methodology}
\label{sec:md}

This section presents DEMVal, a testing method for the error-model semantics of QEC compilers. We first abstract the DEM compilation pipeline that all three subjects instantiate (Section~\ref{sec:pipeline}), then define the symptom-class representation of an error model and the exact differential oracle that compares the emitted model against a reference (Section~\ref{sec:oracle}). Section~\ref{sec:constructions} develops the three reference constructions and the pipeline testing scope each can test.

\subsection{The DEM Compilation Pipeline}\label{sec:pipeline}

\begin{figure}[htbp]
    \centering
    \includegraphics[width=1.0\linewidth]{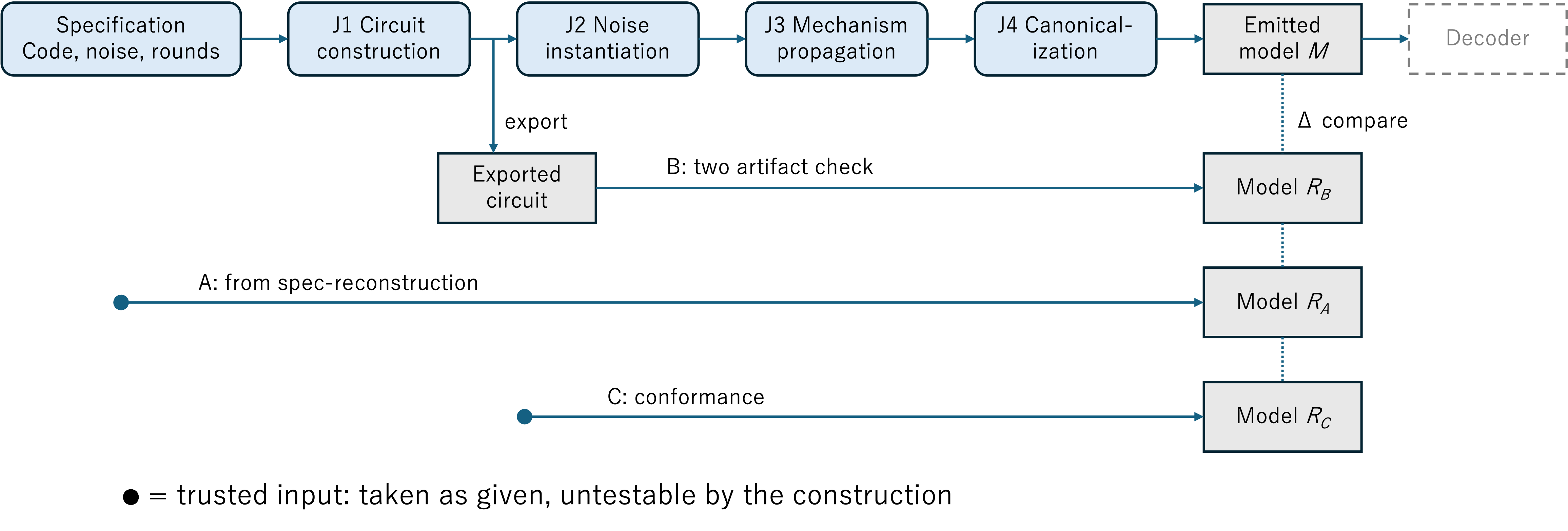}
    \caption{The compile-time pipeline that all three subjects instantiate. Stages J1--J4 build the error model in memory and must construct what the specification means; exporters write the circuit or the model out as artifacts and must encode the object in memory. Each reference construction assumes its trusted input correct and yields a reference model, shown in the right column. The dot marks the trusted input of A and C, and B's is the circuit of stage J1. The harness compares each reference model with the emitted model $M$. The decoder consumes $M$ and lies outside the testing scope.}
    \label{fig:pipeline}
\end{figure}

Figure~\ref{fig:qec_compiler} shows the concrete pipeline. We now abstract its compile-time part into four semantic stages and the artifacts written between them (Figure~\ref{fig:pipeline}), because the abstraction determines where defects can arise and what a reference model can test.

The compile-time part divides into four stages that all three subjects implement. Stage J1 constructs the syndrome-extraction circuit from the specification. Stage J2 instantiates the noise model on that circuit: it places a channel at every site the noise model prescribes, enumerates the error mechanisms of each channel, and converts the channel's parameters into a prior probability for each mechanism. Stage J3 propagates each mechanism through the circuit to determine which detectors and observables it flips. Stage J4 aggregates mechanisms with identical effect and emits the model. The stages carry the same requirement, to construct the object that the specification means, and we call a violation of it a construction defect.

Orthogonal to the stages are the exporters, the code that writes an in-memory object out as an artifact, a file or a foreign object handed to the caller. An exporter must write an artifact that encodes the same object that exists in memory. The artifact itself is data, so it can be wrong but not buggy, and when it is wrong, the defect lies in the exporter. We call such a violation an exporter defect. The two kinds differ in their consequences. A construction defect corrupts the tool's own emitted model whether or not an artifact is written, while an exporter defect leaves the emitted model correct and affects only consumers of the artifact.

The abstraction also fixes what a comparison against a reference model can test. Each construction assumes one input correct and reconstructs its reference from it. Both compared models are built from this trusted input, so a defect in a component between it and either model makes them differ, and a defect in or before it reaches both. The components between the trusted input and the two models form the testing scope.

\subsection{Symptom-Class Models and the Differential Oracle}\label{sec:oracle}

Given the emitted model and a reference, the comparison itself must be defined. Two correct pipelines may write the same error model differently, listing mechanisms in another order, splitting one mechanism into duplicates, or grouping several into one, so a textual comparison would report correct compilations as inconsistent. The oracle compares models in a canonical form that removes this freedom. A symptom class is a pair $s=(D,O)$, the set of detectors and the set of observables that an error mechanism flips. A model is a partial map from symptom classes to probabilities in $(0,1]$. Both the emitted model and the reference are brought into this form by \Gather, which folds mechanisms with the same class into one entry under $x \oplus y = x+y-2xy$, the combination rule of Section~\ref{sec:bg-dem}, and drops the class $(\emptyset,\emptyset)$, whose mechanisms flip nothing the experiment can observe. The result is the decoding problem the model defines.

Canonicalization leaves one further freedom, the assignment of labels. A reference and an emitted model may agree on every mechanism of the same experiment and still share no labels. A symptom class names its detectors and observables by index, and the index is a convention, not part of the specification. If the reference numbers a detector 3 and the tool numbers it 5, every class containing that detector differs between the two models. A reference built from published material cannot always derive the convention a tool adopts. Where this happens, the harness aligns the reference to the emitted model with \textsc{AlignLabels}, which searches a small, declared family of relabelings: the stabilizer relabelings, the choice of representative for the logical observable, and round reversal. The stabilizer relabelings and round reversal permute the detector labels and preserve every probability, so a missing class, an extra class, a merged class, or a deviating rate survives them. The change of representative instead rebuilds the reference for a logical observable that differs from the original by a stabilizer, and the result is another correct reference of the same experiment.

Algorithm~\ref{alg:harness} states the resulting harness. After canonicalization, two models can differ in only two ways, a symptom class present in one model and absent from the other, or a shared class whose probabilities differ, and the comparison computes one quantity for each. The unshared count $u$ is the number of the former, ignoring classes whose probability is below $\epsilon=10^{-15}$ so that rounding noise does not count as structure. The rate deviation $\delta$ is the largest relative difference over the shared classes, $\max_s |M(s)-R(s)|/\max(M(s),R(s))$. The outcome is \Inconsistent{} when $u>0$ or $\delta>\tau$, and \Consistent{} otherwise. The harness records the per-class residuals with the outcome. The tolerance $\tau$ is set per subject. It adopts the tool's own declared comparison tolerance where one exists, and otherwise lies between the numerical noise floor of a correct compilation and the smallest divergence a defect produces.

% ---------- Algorithm 1: shared harness ----------
\begin{algorithm}[htbp]
\caption{DEMVal's shared validation harness for one corpus experiment.}
\label{alg:harness}
\begin{algorithmic}[1]
\Require experiment $c$; compiler under test $T$; construction $\kappa\in\{A,B,C\}$; tolerance $\tau$
\Ensure outcome $\in\{\Consistent,\Inconsistent\}$, per-class residuals
\State $M \gets \Gather\bigl(\textsc{EmitDEM}(T,c)\bigr)$ \Comment{the tool's pipeline}
\State $R \gets \textsc{Reconstruct}_{\kappa}\bigl(c,\ \text{artifacts}(T,c)\bigr)$ \Comment{Alg.~\ref{alg:consA}, \ref{alg:consBC}}
\State $R \gets \textsc{AlignLabels}(R,M)$ \Comment{labels only}
\State $u \gets \bigl|\{\,s \in \dom M \,\triangle\, \dom R : \max(M(s),R(s)) > \epsilon \,\}\bigr|$
\State $\delta \gets \max_{s\in \dom M \cap \dom R} {|M(s)-R(s)|}\,/\,{\max(M(s),R(s))}$
\State \Return $\Inconsistent$ \textbf{if} $u>0 \lor \delta>\tau$ \textbf{else} \Consistent
\end{algorithmic}
\end{algorithm}

The oracle has three properties. (1) It depends only on the two models and the tolerance. (2) It is symmetric, reporting that the testing scope contains a defect without identifying which model is wrong. (3) It is exact, so its sensitivity is set by the tolerance and the numerical noise floor rather than the physical error rate.

\subsection{Reference Constructions}\label{sec:constructions}

The three constructions are the three practical choices of the trusted input, sharing the harness of Section~\ref{sec:oracle} and differing in the input they rely on, which fixes what each can test. 

\subsubsection{A: From-Spec Frame Reconstruction}\label{sec:consA}

Construction A rebuilds the model from the specification alone. The reconstruction proceeds in three steps, recovering the noiseless circuit the specification describes, deriving the definitions of its detectors and observables, and attaching every prescribed error mechanism and propagating it through the circuit to obtain its symptom class. The trusted input is the specification and the channel semantics. The specification $S$ is the source program for deq and the declared parity-check matrices $H_X$ and $H_Z$ with the code's published circuit definition for CUDA-Q QEC. In both cases the channel semantics are the channel's own Pauli decomposition $\{(P_j,q_j)\}_{j\le m}$ and the attachment rule $g$ stating which operations receive noise, every CX, every H, or every measured qubit for CUDA-Q QEC and the noise instructions of the source program for deq, with measurement channels injected before the measurement. Algorithm~\ref{alg:consA} shows the reconstruction. The \textsc{BuildCircuit} step re-executes stage J1 independently, rebuilding the circuit from the specification as the ordered operation list $\mathit{ops}$. The definitions step derives from the specification the detectors and observables $(\mathcal C,\mathcal O)$, each as the set of measurement indices whose parity it reads. Where an element is taken from the tool instead, the \textsc{Deterministic} assertion checks that its parity is constant on the reconstructed noiseless circuit, the defining property of a detector. A definition that fails the check stops the reconstruction, and the experiment ends in an error. The error itself indicates a defect, since either the tool's definitions or the rebuilt circuit must be wrong.

% ---------- Algorithm 2: construction A ----------
\begin{algorithm}[htbp]
\caption{Construction A: from-spec frame reconstruction.}
\label{alg:consA}
\begin{algorithmic}[1]
\Require $S$, $\{(P_j,q_j)\}$, $g$
\Ensure symptom-class model $R$
\State $\mathit{ops} \gets \textsc{BuildCircuit}(S)$ \Comment{independent J1 re-execution}
\State $(\mathcal C,\mathcal O) \gets$ detector and observable definitions from $S$ \Comment{from the tool's model where $S$ declares none}
\State \textbf{assert} \textsc{Deterministic}$(\mathit{ops},\mathcal C,\mathcal O)$
\State $\mathit{sites} \gets \{(t,\vec q) : \mathit{ops}[t] \text{ matched by } g\}$
\ForAll{$e=(t,\vec q,P_j,q_j)$, $(t,\vec q)\in\mathit{sites}$, $j\le m$}
\State inject frame $P_j$ on $\vec q$ at $t$; propagate through $\mathit{ops}$
\State $f_e \in \mathbb F_2^{n_{\mathrm{meas}}} \gets$ measurement flips
\State $s_e \gets \bigl(\{k : \bigoplus_{r\in\mathcal C_k} f_e[r]=1\},\ \{o : \bigoplus_{r\in\mathcal O_o} f_e[r]=1\}\bigr)$
\State $q_e \gets q_j$ \Comment{the mechanism's rate}
\EndFor
\State \Return $\Gather\bigl(\{(s_e,q_e)\}_e\bigr)$
\end{algorithmic}
\end{algorithm}

The remaining step attaches the mechanisms and propagates them, and all of it is independent of the compiler under test. The mechanisms are those of the channel's own Pauli decomposition, one per Pauli at every noise site. Each is an independent event with its own exact prior. Channels that declare the prior directly as a Kraus weight or as the rate of a single Pauli contribute the declared value unchanged. The depolarizing channels of deq instead require a conversion from the exclusive channel parameter to independent priors, so their rates are obtained from Stim's documented implementation of the channel by a tomographic probe~\cite{AAPT}. Each error mechanism $e$, one Pauli error at one noise site, is injected as a Pauli frame and propagated through the reconstructed circuit by Stim. The detector and observable definitions turn its measurement flips $f_e$ into the symptom class $s_e$, which carries the mechanism's own rate $q_e = q_j$, and \Gather{} yields the reference $R$. A's testing scope is the widest of the three, stages J1 through J4, and the exporters lie outside it.

\subsubsection{B: Two-Artifact Internal Consistency}\label{sec:consB}

Construction B checks two artifacts of one compilation against each other, because outputs meant to carry the same semantics must agree. It applies when a compilation emits such a pair, and deq does, exporting the transpiled circuit $C_{\mathrm{stim}}$ as a \texttt{.stim} file together with the compiled binary model $M_{\mathrm{bin}}$, a \texttt{.deq.bin} file. \textsc{Reconstruct}$_B$ in Algorithm~\ref{alg:consBC} takes the detector and observable definitions from the binary model, attaches the corresponding detectors and observables to the exported circuit, and computes the exact model of the result with \textsc{ExactDEM}, Stim's exact model of a noisy circuit. The harness then compares this model against the model read from the binary, whose gathered canonical form plays the role of $M$. B's trusted input is neither artifact but the circuit that stage J1 builds, from which the circuit exporter writes $C_{\mathrm{stim}}$ and stages J2 through J4 compile $M_{\mathrm{bin}}$. Stage J1 lies outside B's testing scope, and the two exporters and stages J2 through J4 lie inside. B is thus the one strictly differential construction of the method.

\subsubsection{C: Published-Spec Conformance}\label{sec:consC}

Construction C checks conformance to a published noise model. It applies when the model is published as an explicit rule set and the tool exports the circuit the noise is placed on. The reference then re-applies the published rules to that circuit and compares the outcome against the model the tool emitted. For Deltakit, the SI1000 and SD6 models publish per-layer rules whose rates are fixed multiples of a single parameter $p$. \textsc{Reconstruct}$_C$ in Algorithm~\ref{alg:consBC} takes the exported noiseless circuit $C_0$ together with the published model $N$. It flattens the circuit into \textsc{tick} layers, where the \textsc{tick} instruction marks the boundary between consecutive time steps. For each layer it classifies the qubits and appends the published noise at the sites the rule prescribes. An operation without a published rule raises an error. The gathered exact model of the re-noised circuit $C_R$ is the reference.

C covers the export edge, together with the placement of noise against the published model. Both lie downstream of the in-memory circuit that the tool exports, which is assumed correct with the artifact because both compared models descend from it. When the export is wrong, the reference is the side that inherits the error. It re-noises the circuit as exported and so differs from the correct model the tool built in memory. Propagation and canonicalization are shared by the two compared models when the tool derives its model with the same semantic engine the reference uses, as for Deltakit, whose emitted model is Stim's model of the noisy circuit it exports, so these stages lie outside C's testing scope. The covered components require exact agreement, so C's tolerance can be set near machine precision.

% ---------- Algorithm 3: constructions B and C ----------
\begin{algorithm}[htbp]
\caption{Reference constructions B and C.}
\label{alg:consBC}
\begin{algorithmic}[1]
\Function{Reconstruct$_B$}{circuit $C_{\mathrm{stim}}$, model $M_{\mathrm{bin}}$}
\State $(\mathcal C,\mathcal O) \gets$ canonical basis of $M_{\mathrm{bin}}$
\State $C' \gets C_{\mathrm{stim}}$ + \textsc{detector}s $\mathcal C$, \textsc{observable}s $\mathcal O$
\State \Return $\Gather\bigl(\textsc{ExactDEM}(C')\bigr)$
\EndFunction
\Statex
\Function{Reconstruct$_C$}{circuit $C_0$, model $N$, rate $p$}
\State $C_R \gets$ empty
\For{each \textsc{tick} layer $L$ of $\textsc{Flatten}(C_0)$}
\State classify $L$'s qubits \Comment{2q pairs, 1q, meas, reset, idle}
\State append $L$, then $N$'s noise at the rule's sites
\EndFor
\State \Return $\Gather\bigl(\textsc{ExactDEM}(C_R)\bigr)$
\EndFunction
\end{algorithmic}
\end{algorithm}

\subsubsection{Testing Scopes and Applicability}\label{sec:testing scopes}

The analysis of Section~\ref{sec:pipeline} turns each construction into a prediction of whether each pipeline component lies inside its testing scope, and Table~\ref{tab:testing scopes} records the predictions. Detection and localization are different requirements, and a testing scope guarantees only the former. The instances differ in how much they assume, and the testing scope shrinks as the trusted input grows. A relies only on the specification and has the widest testing scope, while B and C rely on the compiler's own artifacts and cover less.

Not every construction applies to every subject. B requires one compilation to emit at least two artifacts with the same intended semantics, and among our subjects only deq does. C requires a published rule set and an exported circuit on which to apply it. deq has no such target, because its noise operations are part of the input program rather than the product of a placement stage. CUDA-Q QEC has the placement rule but does not export the circuit at the current API, which assembles the memory circuit internally and returns only the model. C becomes applicable once the circuit is exposed. Construction A requires only the specification and applies to every subject. We implement it for deq and CUDA-Q QEC, and an A instance for Deltakit remains future work. The assignment of constructions to subjects in the evaluation follows this analysis.

\begin{table}[htbp]
\caption{Analytic testing scopes of the four construction instances. \checkmark: inside the testing scope; $\sim$: partially inside; $\times$: outside, for the reason stated; ---: the component does not exist for the subject. The definition content is every construction's input and lies outside every testing scope.}
\label{tab:testing scopes}
\footnotesize
\setlength{\tabcolsep}{4pt}
\begin{tabular}{@{}lllll@{}}
\toprule
Pipeline component & A (deq) & A (CUDA-Q) & B (deq) & C (Deltakit) \\
\midrule
definition content & input & input & input & input \\
J1 circuit synthesis & \checkmark & $\sim$ conventions aligned & $\times$ shared prefix & $\times$ trusted input \\
exporters & $\times$ no artifact & --- & \checkmark & \checkmark \\
J2 noise instantiation & \checkmark & \checkmark & \checkmark & \checkmark \\
J3 propagation & \checkmark & \checkmark & \checkmark & $\times$ shared engine \\
J4 canonicalization & \checkmark & \checkmark & \checkmark & $\times$ shared engine \\
model export & \checkmark & \checkmark & \checkmark & \checkmark \\
downstream of the models & \multicolumn{4}{l}{out of scope for every construction} \\
\bottomrule
\end{tabular}
\end{table}

\section{Experimental Evaluation}
\label{sec:eval}

We evaluate DEMVal on three production QEC compilers, over a detection corpus of 551 memory experiments and a complementary mutation analysis of 136 mutants. Our experiments focus on the following research questions (\textbf{RQ}s):
\begin{itemize}
\item \textbf{RQ1 (Effectiveness and precision).} Does DEMVal detect real semantic defects in production QEC compilers, and does it report only real defects?
\item \textbf{RQ2 (Necessity of exactness).} Could the statistical, sampling-based oracle class have found these defects at realistic operating points?
\item \textbf{RQ3 (Constructions and testing scopes).} How do the three reference constructions differ in what they can test, and do the observed defects confirm the predicted testing scopes?
\end{itemize}

Section~\ref{sec:setup} describes the subjects, the corpora, and the tolerances. The three sections that follow answer the research questions in turn, and Section~\ref{sec:threats} examines threats to validity.

\subsection{Testing Setup}\label{sec:setup}

We evaluate DEMVal on three production QEC compilers. CUDA-Q QEC is tested as cudaq 0.14.2 with cudaq-qec 0.6.0, the most recent PyPI release at the time of the study. Deltakit is tested as release 0.8.0. deq is tested as version 0.3.1 at qdk-ec commit \texttt{bef1be26}, the development version preceding our fix. Following the applicability analysis of Section~\ref{sec:testing scopes}, construction A applies to CUDA-Q QEC, constructions A and B apply to deq, and construction C applies to Deltakit.

The unit of the corpus is a single memory experiment, matching how existing QEC compilers are invoked, as described in Section~\ref{sec:bg-dem}. The grid varies code family, distance, measurement basis, round count, noise channel, and physical error rate as Table~\ref{tab:subjects} lists per subject, plus the gate set for deq. For Deltakit, round counts span both representations of the schedule, the small counts the compiler expresses by unrolling and the larger counts it expresses through \textsc{repeat} blocks. Each excluded combination carries a stated reason, such as a channel no stabilizer of the family can detect or a gate the family lacks. Generation is deterministic, and an ordered manifest indexes the corpus.

Table~\ref{tab:subjects} gives the values varied per subject, 551 experiments in total after the exclusions. For deq we adopt the tool's own declared tolerance of $10^{-6}$, and the other two tolerances lie between the machine-precision agreement of consistent experiments and the smallest defect divergence. The label alignment of Section~\ref{sec:oracle} runs for the CUDA-Q repetition and Steane experiments and returns the identity everywhere except in 27 repetition experiments with two-qubit channels, where it selects round reversal or a reversal of the qubit chain. For the evaluation, each inconsistent experiment is matched to the confirmed defects it exhibits by the shape of its divergence, by a script for CUDA-Q and by the divergence pattern itself for the single defects of deq and Deltakit; the matching is used only to report the results. The mutation analysis uses a separate corpus of mutants, which RQ2 describes.

All reconstruction and comparison stages use Stim 1.16.0 as the semantic engine and run on a desktop workstation with an Intel Core i7-14700KF and 32\,GB of memory. CUDA-Q compilation runs under Linux in WSL, since the compiler targets Linux. The cost of DEMVal itself is small compared to the compilers it tests. The full deq pipeline over 227 experiments completes in 19 minutes end to end, of which the tool's own transpilation and compilation account for 18 minutes, while our frame reconstruction with basis validation takes under 30 seconds and both comparisons together take seconds. On CUDA-Q QEC, reconstructing and comparing all 252 experiments takes 3.5 minutes. The full Deltakit pipeline, including compilation, completes all 72 experiments in 14 seconds. The full mutation analysis of 136 mutants under four oracles runs in under an hour, and the equivalence-checker calls total under 2 minutes. Mutation sites are drawn with a fixed seed.

\begin{table}[htbp]
\caption{Subjects, corpora, and per-subject tolerances. Exp.\ is the number of experiments, and combinations that are not instantiable are excluded with stated reasons. The channels are single- and two-qubit depolarizing and bit-flip errors on the gates and before measurement, phase-flip errors on CUDA-Q QEC, and correlated two-qubit $XX$, $XZ$, and $ZZ$ errors on deq.}
\label{tab:subjects}
\footnotesize
\setlength{\tabcolsep}{4pt}
\begin{tabular}{@{}lp{2.6cm}cp{5.8cm}rl@{}}
\toprule
Subject & Version under test & Constr. & Corpus & Exp. & $\tau$ \\
\midrule
CUDA-Q QEC & cudaq 0.14.2, cudaq-qec 0.6.0 & A & surface $d\in\{3,5\}$, repetition $d\in\{3,5,7\}$, Steane; rounds $\{1,2,3,5\}$; 8 channels; $p=10^{-3}$ & 252 & $10^{-9}$ \\
\addlinespace
deq & 0.3.1, qdk-ec \texttt{bef1be26} & A, B & repetition $d{=}7$, surface $d{=}5$, $[[4,2,2]]$, Bacon--Shor, Floquet; rounds $\{1,2,3,5\}$; 6 channels; rates $10^{-3}$ to $0.15$; 2 gate sets & 227 & $10^{-6}$ \\
\addlinespace
Deltakit & 0.8.0 & C & rotated planar $d\in\{3,5\}$, unrotated $d{=}3$; rounds 1--6; SI1000, SD6; $p=10^{-3}$ & 72 & $10^{-12}$ \\
\bottomrule
\end{tabular}
\end{table}

\subsection{RQ1: Effectiveness and Precision}

DEMVal reported divergences in all three subjects, and inspecting the inconsistent experiments yielded six distinct defects, summarized in Table~\ref{tab:defects}. We reported all six to the developers. The developers confirmed every report, and fixes have been accepted in all three projects, including our own patch for the deq defect. Three of the six defects yield models that are wrong under any noise semantics, since entire symptom classes are absent from or merged into the emitted model, and all six were present in release versions.

On CUDA-Q QEC, 187 of 252 experiments are inconsistent, and 65 are consistent. On deq, both the from-spec reconstruction and the internal two-artifact check report the same 93 of 227 experiments as inconsistent, and this set coincides with the experiments whose noise contains a depolarizing channel, which is the channel family affected by the conversion defect. On Deltakit, 36 of 72 experiments are inconsistent, and this set coincides with the round counts the compiler expresses through a \textsc{repeat} block, which is the representation affected by the export defect. The divergences match their defects quantitatively. The deq deviation is approximately $p/3$ for single-qubit and $7p/15$ for two-qubit depolarizing channels at every rate, more than two orders of magnitude above the comparison tolerance, and the CUDA-Q rate residual equals $p/3$ on the single-qubit depolarizing experiments, the value that additive instead of exclusive composition produces.

DEMVal reports no false alarms on this corpus. Consistent experiments match the reference at machine precision, with relative deviations at or below $10^{-15}$ on shared classes, including all 134 consistent deq experiments. The two independent deq checks agree on all 227 outcomes, and the determinism validation of the derived detectors and observables passes on every experiment.

\begin{table}[htbp]
\caption{The six defects DEMVal detected. All six were confirmed by the developers and fixed, the deq fix being our own patch. Issue prefixes identify the subjects, where qdk-ec is deq's repository, and cudaqx is CUDA-Q QEC's. Structural divergences denote symptom classes absent from or merged in the emitted model; rate divergences are relative deviations on shared classes, against the per-subject tolerances of Table~\ref{tab:subjects}.}
\label{tab:defects}
\footnotesize
\begin{tabular}{llll}
\toprule
Issue & Pipeline Component & Defect & Divergence \\
\midrule
qdk-ec\#68 & noise conversion & inexact channel-to-mechanism rate conversion & up to $5.3\times10^{-2}$ \\
cudaqx\#606 & enumeration & two-qubit channels attached to only the first CX operand & structural \\
cudaqx\#607-CQ2 & canonicalization & same-syndrome, different-observable columns merged & structural \\
cudaqx\#607-CQ3 & canonicalization & zero-syndrome logical columns dropped & structural \\
cudaqx\#607-CQ4 & canonicalization & duplicate rates composed additively & $3.3\times10^{-4}$ \\
deltakit\#281 & circuit export & \textsc{repeat} export drops a boundary \textsc{tick} & up to $1.7\times10^{-1}$ \\
\bottomrule
\end{tabular}
\end{table}

\begin{tcolorbox}
\textbf{Answer to RQ1: DEMVal detected six previously unknown semantic defects across all three compilers, and every one was confirmed by the developers and fixed. Its reports are precise: every inconsistent experiment exhibits one of the six confirmed defects, and the consistent experiments match the reference at machine precision. }
\end{tcolorbox}

\subsection{RQ2: Exact Oracle versus Statistical Oracles}

To answer RQ2, we compare the statistical oracle class that existing quantum software testing uses with our exact model comparison on mutants and on the six real defects. We start from a mutation analysis of the compiled model. Nine mutation operators mutate deq's compiled model on three targets, two repetition-code targets at elevated rates of $0.05$ and $0.15$ and one surface-code target at the realistic $p=10^{-3}$, and span the range from the replayed conversion defect through dropped and duplicated mechanisms, altered symptom sets, and rate perturbations to a moment-preserving correlation shift and mass structural loss. A mutant is live when it changes the canonical model, which is decided by an oracle-independent comparison, and all four oracles are scored on the same live mutants. We evaluate two exact and two statistical oracles. The two exact oracles are our model comparison and deq's own equivalence checker. The two statistical oracles are built on the first and second moments of a sample, the flip rate of each detector or observable and the parity flip rate of each pair of them, which are the simplest statistics a sample exposes. For each we take the strongest such test, a $5\sigma$ $z$-test on the detector, observable, or pair whose flip rate changes most, so a defect it misses cannot be detected by any test that observes the same statistic. The probability that this test detects a mutant at a given number of samples is computed analytically. 

\begin{table}[htbp]
\caption{Detected live mutants per family on the realistic surface-code target ($p=10^{-3}$). ProgEq is deq's equivalence checker at its $10^{-6}$ tolerance. DEM is our exact comparison, at the two tolerances shown. The last two columns are the optimal $5\sigma$ $z$-tests on first and second moments at $10^{7}$ samples. The exact oracles detect every mutant except the rate-perturbation mutants built below the tolerance.}
\label{tab:power}
\footnotesize
\setlength{\tabcolsep}{4pt}
\begin{tabular}{lccccc}
\toprule
Mutation operator & ProgEq & DEM@$10^{-6}$ & DEM@$10^{-3}$ & 1st mom.\ @$10^{7}$ & 2nd mom.\ @$10^{7}$ \\
\midrule
F1 conversion defect (interpolated replay) & 5/5 & 5/5 & 0/5 & 0/5 & 0/5 \\
F2 additive merge & 2/2 & 2/2 & 0/2 & 0/2 & 0/2 \\
F3 dropped mechanism & 6/6 & 6/6 & 6/6 & 0/6 & 0/6 \\
F4 duplicated mechanism & 6/6 & 6/6 & 6/6 & 0/6 & 0/6 \\
F5a symptom dropped & 6/6 & 6/6 & 6/6 & 0/6 & 0/6 \\
F5b symptom added & 6/6 & 6/6 & 6/6 & 0/6 & 0/6 \\
F6 rate perturbation ($\varepsilon$ probe) & 4/7 & 4/7 & 1/7 & 0/7 & 0/7 \\
F7 readout flip & \multicolumn{5}{c}{no live mutants (all three sites dead)} \\
F8 moment-preserving shift & 1/1 & 1/1 & 1/1 & 0/1 & 1/1 \\
F9 mass drop ($f{=}0.01$--$0.6$) & 5/5 & 5/5 & 5/5 & 5/5 & 5/5 \\
\midrule
Total & 41/44 & 41/44 & 31/44 & 5/44 & 6/44 \\
\bottomrule
\end{tabular}
\end{table}

Table~\ref{tab:power} gives the results on the realistic target. The two exact oracles catch all 41 live mutants whose divergence lies above their tolerance and none of the three rate perturbations that lie below it, while the two moment tests catch only 5 and 6 mutants. Outside the mass-drop family, at $p=10^{-3}$ the optimal first-moment test detects none of the 39 live mutants even at $10^{7}$ samples, and the second-moment test detects one. The clearest case is a dropped mechanism. Removing an entire two-qubit depolarizing mechanism shifts the emitted model by up to $50\%$ relative, and both exact oracles detect it on every site, yet its largest effect on any per-detector flip rate is only $6.6\times10^{-5}$, so the optimal test detects it with probability at most $0.004$ at $10^{7}$ samples. Empirical sampling at $10^{6}$ shots confirms the analytic prediction, with worst $z$ scores between $1.4$ and $2.1$ over three seeds, indistinguishable from sampling noise. The one mutant the second moment catches is the moment-preserving shift, whose correlation change preserves every per-detector flip rate, so the first-moment test does not detect it at any number of samples, while the second-moment test does. Exact comparison subsumes this hierarchy and detects every instance above its tolerance.

The cases so far show individual mutants that sampling does not detect, but not where its threshold lies. To locate it, the ninth operator varies the absolute change directly, dropping a variable fraction of the compiled model's mechanisms. On the realistic target even 1\% of the mechanisms, 15 of 1440, changes a per-detector flip rate by $1.1\times10^{-3}$ and is detected by the first-moment test already at $10^{6}$ samples. The optimal test starts to detect near an absolute change of $10^{-4}$, between a single dropped mechanism and 1\% of the model. This boundary classifies the six real defects, which split across it. The two rate-level defects fall below it, changing no per-detector flip rate by more than about $10^{-5}$, and are detected with probability $0.000$ even at $10^{7}$ samples. The four mechanism-level defects, the merged, dropped, and missing classes and the misplaced export noise, lie above it and change detector or observable flip rates by $3\times10^{-4}$ to $3.4\times10^{-2}$. The three CUDA-Q defects are detected with probability near one at $10^{4}$ to $10^{5}$ samples; the export defect is detected at $10^{6}$ samples under SD6, and under SI1000 only in part even at $10^{7}$ samples, in 8 of 18 experiments above $0.9$. For deltakit\#281, the only defect whose faulty circuit is available to sample, sampling at $10^{6}$ shots confirms the analytically computed probability. Detection by sampling is therefore set by the absolute change in the syndrome statistics, not by the relative size of the model error. The dropped mechanism shifts the model by 50\% relative and remains undetected, while the export defect shifts it by less than 2\% relative under its lower-noise model and is still detected at $10^{6}$ to $10^{7}$ samples. What sampling leaves undetected at this operating point is the rate-level class whose consequence is a decoder calibrated against wrong rates.

\begin{tcolorbox}
\textbf{Answer to RQ2: Sampling could have found only the four mechanism-level defects, and all four only with millions of samples. The two rate-level defects lie beyond even the analytically optimal moment tests. Exact model comparison detects all six with no samples, so the oracle must be exact rather than statistical.}
\end{tcolorbox}

\subsection{RQ3: Constructions and Testing Scopes}

The testing scopes are predictions, and the six defects are the evidence. Table~\ref{tab:rq3} places each defect at its pipeline component and the construction that detected it. The six defects lie inside the testing scopes of their detecting constructions, none in J1 or J3, and no single construction detects all six.

\begin{table}[htbp]
\caption{The six defects against the predicted testing scopes, by pipeline component and detecting construction.}
\label{tab:rq3}
\footnotesize
\setlength{\tabcolsep}{4pt}
\begin{tabular}{@{}lll@{}}
\toprule
Defect & Component & Detected by \\
\midrule
conversion (qdk-ec\#68) & J2 & A, B \\
enumeration (cudaqx\#606) & J2 & A \\
canonicalization $\times$3 (cudaqx\#607) & J4 & A \\
export (deltakit\#281) & exporter & C \\
\bottomrule
\end{tabular}
\end{table}

The deq conversion defect lies where the testing scopes of A and B overlap. A and B report the same 93 experiments as inconsistent but localize differently. A, whose reference is derived from the specification alone, shows a pure rate deviation that points to the parameter conversion, while B, whose inputs are the artifacts themselves, reports only that the two artifacts disagree. The two constructions' per-experiment agreement is the redundancy their overlap predicts.

The exporter defect separates the two defect kinds in practice. Deltakit builds its circuit correctly in memory, and the model derived from that circuit is correct. The exporter drops a \textsc{tick} at every \textsc{repeat} boundary, so the artifact carries an incorrect layer structure, and the maintainers fixed the defect in the exporter. Construction C rebuilds its reference from the artifact, so the export error appears as a divergence between the rebuilt model and the correct model the tool holds in memory. This is the case in which the symmetry of the outcome matters: the reference, which follows the artifact, is the side that carries the error, and the emitted model is the correct one. The defect is nevertheless the tool's, because its exporter produced the artifact the reference followed. Of the three constructions only C covers this defect class, since A reads no artifact and B needs a second artifact, which Deltakit does not emit.

The defects concentrate in J2, J4, and an exporter, the components where a tool performs its own probability arithmetic, column canonicalization, and serialization. The prediction holds for every component: each defect lies inside the testing scope of its detecting construction, and no single construction covers all six defects.

\begin{tcolorbox}
\textbf{Answer to RQ3: The six defects fall where the constructions' testing scopes predict, and no single construction covers all of them. The choice of construction, fixed by the input it relies on, is therefore a design dimension of the method, not an implementation detail.}
\end{tcolorbox}

\subsection{Threats to Validity}\label{sec:threats}

The first threat is a defect in the reference itself, which could report correct compilations as inconsistent or agree with wrong ones. Three observations bound it. Consistent experiments match the emitted model at machine precision, $10^{-15}$ or below, so a wrong reconstruction would have to reproduce the tool's model. On deq, two constructions with different trusted inputs agree on every outcome, and the determinism validation passes on every experiment. And every reported divergence was confirmed by the developers as a tool-side defect. 

The second threat is the per-subject tolerances, which could be tuned to produce the outcomes. In practice every outcome is unchanged across more than eight orders of magnitude of tolerance, since consistent experiments sit at machine precision while real divergences lie far above any plausible tolerance, and 178 of the 316 divergences are structural, which no tolerance affects.

The third threat is the matching of inconsistent experiments to the six defects: a further defect producing the same kind of divergence would be grouped with a known one, so six is a lower bound on the defects present.

\section{Related Work}
\label{sec:rw}

This section relates DEMVal to five adjacent lines of work: testing of quantum software stacks, which uses the statistical oracle class evaluated in RQ2; classical compiler testing and translation validation, which our harness follows; formal verification of quantum compilation and of error correction; prior work at the detector-error-model layer itself; and the evaluation of decoders and protocols downstream of that layer.

\paragraph{Testing quantum software stacks.}
Differential and metamorphic testing have been adapted to quantum computing platforms. QDiff~\cite{QDiff} generates semantics-preserving variants of quantum programs and compares stacks through sampled output distributions judged by Kolmogorov--Smirnov and cross-entropy statistics. MorphQ~\cite{MorphQ} pairs a program generator with quantum-specific metamorphic relations and reveals mostly crash bugs. QEMI~\cite{QEMI} transfers equivalence modulo inputs to quantum stacks. Fuzzing has also been directed at quantum simulators~\cite{FuzzSimu}, and Fuzz4All~\cite{Fuzz4All} generates inputs with large language models against crash-class oracles, with Qiskit among its targets. An empirical study of 223 platform bugs~\cite{BugsEmp} found that roughly 40\% are quantum-specific and that they concentrate in the components that represent, compile, and optimize quantum abstractions. Our subjects belong to that layer. All of these approaches apply to general quantum circuits and check correctness by crashes or by comparing sampled behavior, as do statistical assertions at the program level~\cite{StaAss}. None checks the semantics of the compiled error model itself. Our oracle is exact, not sampled, and RQ2 quantifies the consequence of this difference.

\paragraph{Compiler testing and translation validation.}
Our harness builds on differential compiler testing~\cite{McKeeman1998DifferentialTF, BugsinCCompiler, EMI}. Classical differential testing lets multiple comparable implementations serve as one another's oracle, but that approach does not apply here, because a detector error model depends on the tool's own syndrome-extraction circuit, and two compilers correctly produce inequivalent models for the same code and noise. DEMVal therefore constructs the reference itself, as a pseudo-oracle~\cite{pseudo-oracle}. Compilations are validated individually against an independently derived reference, as in translation validation~\cite{TV, TVOC, Alive2}, rather than the compiler being verified once and for all~\cite{FVofCompCert}. Construction B is the exception, since there the two artifacts of one compilation check each other, with no input from outside the compilation. The distinction in our pipeline abstraction between construction defects in stages and exporter defects corresponds to the classical separation between wrong-code bugs and bugs in the code that writes a compiler's output, which verified compilers leave outside their core proofs.

\paragraph{Verified quantum compilation and equivalence checking.}
VOQC~\cite{VOQC} and Giallar~\cite{Giallar} prove that the optimizer and transpiler passes preserve the semantics of the circuit. QCEC decides the equivalence of two circuit descriptions across a compilation flow~\cite{QCEC, QCECVerify}. Formal methods have also been applied to error correction itself. Veri-QEC~\cite{Veri-QEC} proves assertions about fault-tolerant protocol programs in a dedicated program logic, and an end-to-end Lean formalization of the stabilizer stack~\cite{Lean-QEC} certifies code-theoretic properties such as distance inside a proof kernel. These guarantees attach to verified reimplementations, to programs written in the respective logics, or to the underlying theory. The noise instantiation and canonicalization layer that produces the detector error model lies outside their verified scope, and it is where all six defects that we report arise. Moreover, our subjects are the released production implementations themselves. deq in particular builds its implementation on the data structures and algorithms of stabilizer circuit verification~\cite{StabCirVeri}. Our results show that the artifact layers around this theoretically grounded core, including channel parameter conversion and serialization, still miscompile. Independent testing remains valuable even for toolchains built directly on verification theory.

\paragraph{The detector-error-model layer.}
Stim~\cite{stim} defines the detector error model representation and serves as our trusted semantic engine, both for frame propagation and for the tomographic probe that provides the mechanism rates. Construction C checks conformance against published circuit noise models~\cite{HoneycombMemory}. Two recent lines are adjacent to ours but address different problems. Quasilinear equivalence checking for detector error models~\cite{EqCheck4DEM} decides exact equivalence of two given models, a relation that even correct models from independent implementations do not satisfy, since their labels and rounded rates differ. Our contribution is complementary. We construct the reference model that a compiler's output must be compared against, and we report the divergences that appear. Reconstruction of detector error models~\cite{ReconDEM} infers a noise model from experimentally measured syndrome statistics for decoding. The information flows in the opposite direction from ours, from data to model rather than from specification to model.

\paragraph{Evaluating decoders and error correction protocols.}
A substantial body of work targets the layer downstream of the DEM. Scalable QEC testing~\cite{ScalableTestingofQEC} estimates logical error rates at large distances through stratified fault injection. Recent work replaces Monte Carlo estimation with a systematic analysis of decoder accuracy~\cite{AnalyzingDecoders}, which, like our exact oracles at the compiler layer, avoids sampling. The decoder implementations these efforts evaluate, such as PyMatching~\cite{PyMatching, SparseBlossom}, compute exact minimum-weight matchings in their algorithmic core. All of these lines focus on how well a decoder or protocol performs against a given error model. None of them checks whether the error model provided to the decoder is the one the compiler should have produced. Our comparison checks that, before the model reaches the decoder.

\section{Conclusion}
\label{sec:con}

We presented DEMVal, a testing method for the error-model semantics of QEC compilers. DEMVal abstracts the compile-time pipeline into four stages and the artifacts written between them, reconstructs a reference model whose trusted input and testing scope are explicit, and compares the reference against the emitted model exactly at the level of symptom classes. On CUDA-Q QEC, Deltakit, and deq the method detected six silent defects, all confirmed by the developers and fixed, with no false alarm over 551 experiments. The mutation analysis determined the boundary of what sampling can detect, and the rate-level defects lie beyond it at realistic error rates, while the placement of the six defects confirmed the testing scopes the constructions predict.

Two directions follow from this work. As further toolchains expose the inputs the method requires, the method applies beyond the three compilers we study. A second direction lies downstream of the emitted model, in the decoder's use of it, which current testing does not examine.

% \begin{acks}
% \end{acks}

\bibliographystyle{ACM-Reference-Format}
\bibliography{sample-base}

% \appendix

\end{document}